\documentclass[preprint,30pt,usegeometry]{elsarticle}
\usepackage[left=2cm,top=2cm,right=2cm,bottom=2 cm,bindingoffset=1.cm]{geometry}

\usepackage{amssymb}

\usepackage{graphicx}
\usepackage{dcolumn}
\usepackage{bm}

\usepackage[utf8]{inputenc}
\usepackage[T1]{fontenc}
\usepackage{mathptmx}
\usepackage{etoolbox}
\usepackage{amsmath}
\usepackage{amssymb}
\usepackage{hyperref}
\usepackage{verbatim}
\usepackage{algorithm2e}
\usepackage{lmodern}

\usepackage{graphicx}
\usepackage{tikz}
\usetikzlibrary{shapes.geometric, arrows}
\usepackage{booktabs}

\newcounter{bla}

\journal{Computer Physics Communications}

\newcommand{\vbm}[1]{\boldsymbol{#1}}

\begin{document}

\begin{frontmatter}



\title{GREENY: A Full-F 2D Gyrofluid Reconnection Code}


\author[1]{F. F. Locker}
\author[1,2]{ M. Held}
\author[1]{T. M. Stocker-Waldhuber}
\author[1]{A. Stürz}
\author[1]{M. Rinner}
\author[1]{A. Kendl}

\affiliation[1]{Department of Ion Physics and Applied Physics, University of Innsbruck, Technikerstrasse 25, 6020 Innsbruck, Austria, franz.locker@uibk.ac.at}
\affiliation[2]{Department of Mathematics and Statistics, UiT The Arctic University of Norway, 9037 Tromsø , Norway}
\cortext[author] {F. F. Locker.\\\textit{E-mail address:} franz.locker@uibk.ac.at}

\begin{abstract}
We present the 2D gyrofluid magnetic reconnection code GREENY (Gyrofluid Reconnection with Extended Electromagnetic Nonlinearity). After a brief introduction to gyrofluids, magnetic reconnection, and the implemented models, we discuss the numerical framework and the algorithmic treatment of the quasi-neutrality condition and Amperè's law. Next, we present solver tests, conservation laws, and the influence of artificial subgrid dissipation on Harris-sheet magnetic reconnection. Finally, we show different applications, initial conditions and present example simulations.
\end{abstract}


\end{frontmatter}



{\bf PROGRAM SUMMARY}

\begin{small}
\noindent
{\em Program Title:GREENY}                                          \\
{\em CPC Library link to program files:} (to be added by Technical Editor) \\
{\em Developer's repository link:} https://git.uibk.ac.at/c7441315/greeny \\
{\em Code Ocean capsule:} (to be added by Technical Editor)\\
{\em Licensing provisions:} MIT \\
{\em Programming language:}         C/C++                          \\
{\em Supplementary material:} Granalysis, \begin{verbatim} vortex_experiments.pdf, gyromod_derivation.pdf \end{verbatim}  \\
{\em Nature of problem(approx. 50-250 words):}\\
Solves 2D isothermal electromagnetic gyrofluid equations of magnetic reconnection with self-consistent finite Larmor radius effects. The simulations can be done, using full-F, $\delta$F models with arbitrary wavelength polarisation or long-wavelength limit. To invert Ampère's law, multiple solvers, with and without Boussinesq-Oberbeck approximation, are available.\\
{\em Solution method(approx. 50-250 words):}\\
Finite difference solver for the dynamical gyrofluid density and momentum equations (Adams-Bashforth scheme, Arakawa scheme) with spectral and iterative solvers for evaluation of the gyrofluid polarisation equation, gyroaveraging operators and Ampère's law.\\
{\em Additional comments including restrictions and unusual features (approx. 50-250 words):}\\
Requires OpenMP, FFTW3 and NetCDF
   \\

\end{small}

\section{\label{sec:level1}Introduction \protect\\}
In magnetized plasmas, magnetic reconnection (MR) is a mechanism to transfer magnetic field energy into kinetic energy and is hence of high importance in numerous systems like stellar coronae, the earths magnetotail and magnetically confined fusion. Its essence is the breakdown of ideal MHD and the resulting violation of flux conservation \citep{Boozer}. Since the investigation of magnetic reconnection started with Sweet, Parker (Sweet 1958; Parker 1957) and Petschek, the research focus widened from collisionally resistive MR to its collisionless counterpart \citep{Parker1,Parker2,Petschek,sweet,Uzdensky}. Spanning the numerous problems (e.g. \textit{interplay of scales} and the \textit{trigger problem}), many publications sum up the current state of research \cite{Jafari,Parker,Zweibel,MagRec}.

 In the context of magnetically confined nuclear fusion, magnetic reconnection is encountered via a poloidal tearing of the magnetic field and results in island formation, leading to a radial short cut and hence change in radial transport. This flattening of the radial profile is referred to as the sawtooth crash \cite{Letsch}. In low $\beta=\mu_0 p/B^2 \ll 1$ and potentially large aspect ratio scenarios, a resonant flux surface becomes tearing unstable, forming multiple X-points and islands. This can happen on large scales via a poloidal mode in the magnetic field, leading to turbulence (top-down), or (bottom-up) when small seed-islands are formed by turbulence, leading to neoclassical tearing modes \cite{Agullo,Sauter}. Without heat flux, such islands normally saturate at a given size, and, including neoclassical effects, must exceed a threshold size to grow \cite{Fitzpatrick,Rutherford}. 
It follows that even for a tearing stable configuration, one has to investigate the turbulent formation and growth of seed islands. The many ways this can happen include the formation of plasmoids for elongated, tearing unstable current sheets, with the possibility of additional turbulence due to Kelvin-Helmholtz instability \citep{Granier,Borgogno}. The formation of plasmoids, commonly referred to as the plasmoid instability, increases the reconnection rate due to the coexistence of multiple reconnection zones \citep{Comisso2}. Although finite Larmor radius effects (FLR) play a minor role in scenarios with small stability parameter $\Delta$, it has also been shown, that electron inertia and finite FLR effects can lead to fast magnetic reconnection in a suitable environment \cite{Ottavini,Grasso,Comisso, Biancalani}.
This work aims to present the Full-F 2D gyrofluid simulation code GREENY (Gyrofluid Reconnection with Extended Electromagnetic Nonlinearity) with the purpose of investigating collisionless magnetic reconnection in a typical parameter-space of nuclear fusion plasmas. In this context, collisionless means, that parallel resistive terms are neglected in the equations and the reconnection is driven by inertia and numerical/hyper viscosity only. GREENY was developed with TIFF (Gyrofluid turbulence in full-f and full-k) as reference and uses similar solvers for the polarisation equation and time/spatial discretization \cite{Kendl}. Although there is some algorithmic overlap, the routines were tailored to the problems of magnetic reconnection while Hasegawa-Wakatani drift wave dynamics are only built in for code verification and cross checking. While arbitrary wavelengths can be considered in the polarisation equation,  Full-F translates to a time evolution of the total densities. The resulting model is referred to as Full-F non-Oberbeck-Boussinesq (NOB) model with arbitrary wavelength polarisation, where in the presented work the gyro radius is assumed constant $\rho_s=\sqrt{T_s m_s}/(eB)=const$. The magnetic background field $B=B_0=const.$ is in that sense local, which can be interpreted as a partial OB-approximation. As a consequence, only perturbations in the magnetic field are considered and all model variations are isothermal \cite{Held}. In addition to that, the \(\delta F\) limit of the model splits the densities in a constant Background density $N_0$ and perturbations $\delta N_z$, resulting in a \(\delta F\) set of a equations, including an OB-approximated arbitrary wavelength polarisation equation \ref{polapprox}. The difference to TIFF is in the inclusion of the self-consistent calculation of fluctuations in the perpendicular magnetic field via Ampère's law. The algorithm for this is sketched out in section \ref{fieldeq} and includes further assumptions, depending on the model and initial condition. In the general gyrofluid picture, GREENY has the possibility of treating both Full-F and \(\delta F\) magnetic reconnection with finite Larmor radius effects for an ion-/electron temperature ratio $\tau_i=T_i/T_e \neq 0$.  While \(\delta F\) gyrofluid models are based on partly linearized \(\delta F\) gyrokinetics, Full-F gyrofluid models are obtained by taking moments of the Full-F gyrokinetic Vlasov equation \cite{Madsen,ScottV2}. The complexity lies in the treatment of the quasi-neutrality constraint and Ampère's law, where additional approximations have to be applied to find invertible forms. These approximations have to be chosen carefully, and different options are explained in section \ref{model}. 
The paper is organized as follows. In section \ref{model}, we introduce the \(\delta F\) and Full-F model equations. Then, in section \ref{num}, we explain the numerical implementation and present some verification and validation tests as well as some remarks on choosing appropriate sub grid-dissipation. In section \ref{inibound} we explain the different initial and boundary conditions and the control parameters set via the input file. Finally, in section \ref{examples} and \ref{conclu}, we show example simulations and give an outlook on the further development of GREENY.

 \section{Model Equations}
 \label{model}
\subsection{Full-F collisionless MR model}
The collisionless magnetic reconnection model is derived from the Full-F gyrofluid model of Madsen \cite{Madsen}.  For the purpose of this paper we take the low beta $\beta \ll 1$ limit of the Full-F model, and neglect all curvature effects, also assuming a strong guide magnetic field $B_0$. The evolution equations are normalized using the Bohm normalisation scheme \ref{tab:norm}
\begin{table}[htp]
    \centering
    \begin{tabular}{|l|l|}
    \hline
       time  & $ t:=\frac{c_s}{\hat{L}_{\hat{\perp}}}\hat{t}$\\
       \hline
       gyrocenter number density  & $N_z:=N_0^{-1} \hat{N}_z$\\
       \hline
       poisson bracket  & $ \{\cdot, \cdot \}_{\perp}:=\rho_{s,0}^{-2}[\cdot,\cdot]_{\hat{\perp}}$\\
       \hline
        mag. field magnitude  & $B:=B_0^{-1} \hat{B}$\\
       \hline
         gen.   electric  potential  & $\psi_z:=\frac{T_e}{e}\hat{\psi}$\\
       \hline
       par. velocity  & $U_z:=\frac{1}{c_s}\hat{U}_z$\\
       \hline
        par.  mag. vectorpotential & $A_{\hat{\parallel}}:=(\beta B_0 \rho_{s})^{-1} \hat{A}_{\hat{\parallel}}$\\
       \hline
        perp.  nabla  operator  & $\nabla_\perp:=\rho_{s,0}\hat{\nabla}_{\hat{\perp}}$\\
       \hline
       \end{tabular}
    \label{tab:norm} \caption{Bohm normalisation scheme of the dependent fields.}
\end{table}

with the system constants and parameters \ref{tab:param}
\begin{table}[h]
    \centering
    \begin{tabular}{|l|l|}
    \hline
      species temp.  & $ T_z$ \\
       \hline
       species mass  & $m_z$\\
              \hline
        charge number  & $Z_z$\\
       \hline
           charge  & $q_z:=eZ_z$\\
       \hline
               drift scale  & $\rho_{s}:=\sqrt{T_e m_i}/(eB_0)$\\
       \hline
       \hline
       temp. ratio  & $\tau_z:=\frac{T_z}{Z_z T_e}$\\
       \hline
       mass ratio& $\mu_z:=\frac{m_z}{Z_z m_i}$\\
       \hline
         electron  beta  & $\beta=\mu_0 N_{e0} T_e/B_0^2$\\
                \hline
       sound  velocity  & $c_{s}:=\sqrt{T_e/m_i}$\\
          \hline
        $ \quad E \times B$ velocity & $|\vbm{U_{E}}|=1/B_0| \nabla_{\perp} \hat{\psi}_z|$\\
                   \hline
         perp. poisson bracket & $\begin{aligned} 1/B_0 (\vbm{\hat{b}} \times \nabla_{\perp} f ) \cdot \nabla g =\\
         =1/B_0 [f,g]_{\perp}\end{aligned}$\\
       \hline
       \end{tabular}
    \label{tab:param} \caption{Physical quantities, system parameters and identities.}
\end{table}
where we split the magnetic field unit vector $\vbm{\bar{b}}=\vbm{\hat{b}}+\vbm{\tilde{b}}$ in the background $\vbm{\hat{b}}$ and fluctuating $\vbm{\tilde{b}}=\frac{1}{B_0}\nabla A_{\parallel} \times \vbm{\hat{b}}$ part. The parallel gradient of a field $\hat{f}$ accounts via
\begin{align}
\hat{\bar{\nabla}}_{\hat{\parallel}} \hat{f}:=\hat{\nabla} _{\hat{\parallel}} \hat{f}+\vbm{\tilde{b}}\cdot \hat{\nabla} \hat{f} = \hat{\nabla}_{\hat{\parallel}} \hat{f}-\frac{\beta}{\delta \hat{B}}[\hat{A}_{\hat{\parallel}},\hat{f}]_{\hat{\perp}}
\end{align}
the fluctuating part of the magnetic field for the magnetic flutter terms in the models. In this normalisation, dimensionless $\delta = \rho_s/\hat{L}_{\hat{\perp}}$ appears as a scaling parameter for the poisson brackets and hence the dynamics of the simulation. However, since the characteristic gradient length will be in the order of $\rho_s$, we set $\delta=1$. The gyrocenter $E \times B$-drift consists of the gyroaverage and polarisation part of the gyrofluid potential $\hat{\psi}_z=\hat{\psi}_{1,z}+\hat{\psi}_{2,z}$ \cite{Held3}, 
\begin{align}
\hat{\psi}_{1,z} :=\hat{\Gamma}_{1,z} \hat{\phi}, \\
\hat{\psi}_{2,z}:= - \frac{\mu_z}{2} |\hat{\nabla}_{\hat{\perp}}\hat{\Gamma}_{1,z} \hat{\phi} |^2.
\end{align}
The exact form of the terms depends on the model assumptions discussed in  table \ref{polapprox}.
For the rest of the paper, $\beta$ is absorbed into $\hat{A}_{\hat{\parallel}}$. The normalized gyroaverage and screening operators, in their Padé-approximated form, are~\citep{Held2}

\begin{align}
\hat{\Gamma}_{0,i}&:=\frac{1}{1-\tau_i \mu_i \hat{\Delta}_{\hat{\perp}}},& \hat{\Gamma}_{1,i}&:= \frac{1}{1-\frac{\tau_i}{2} \mu_i\hat{\Delta}_{\hat{\perp}}},\\
\hat{\Gamma}_{0,e}&:=\frac{1}{1+\mu_e \tau_e \hat{\Delta}_{\hat{\perp}}},& \hat{\Gamma}_{1,e}&:= \frac{1}{1+\frac{\mu_e}{2} \tau_e \hat{\Delta}_{\hat{\perp}}},
\end{align}
and since $|\mu_e| \ll 1$ for e.g. proton/electron plasmas, we can neglect the electron FLR effects e.g. $\hat{\psi}_e=\hat{\phi}$. The normalized and implemented form of the Full-F equations read
%
%
%
\begin{align}
    \frac{\partial}{\partial t} \hat{n}_e &=
    -\left[\hat{\phi},\,\hat{n}_e\right]_{\hat{\perp}} 
    + \left[\hat{A}_{||},\,\hat{n}_e\,\hat{u}_e\right]_{\hat{\perp}},
\label{eq:4.1_ff1}
    \\
    \frac{\partial}{\partial t} \hat{N}_i &=
    -\left[\hat{\psi}_i,\,\hat{N}_i\right]_{\hat{\perp}} + 
    \left[\hat{\Gamma}_1 \hat{A}_{||},\,\hat{N}_i\,\hat{U}_i\right]_{\hat{\perp}},
   \label{eq:4.1_ff2}
    \\
    \frac{\partial}{\partial t}\left(\hat{u}_e+\frac{1}{\mu_e}\,\hat{A}_{||}\right) &=
    -\left[\hat{\phi},\,\hat{u}_e+\frac{1}{\mu_e}\,\hat{A}_{||}\right]_{\hat{\perp}} + \\
    +\left[\hat{A}_{||},\,\frac{1}{2}\,\hat{u}_e^2\right]_{\hat{\perp}}
    & -\frac{1}{\mu_e}\,\left[\hat{A}_{||},\,\ln(\hat{n}_e)\right]_{\hat{\perp}},
\label{eq:4.1_ff3}
    \\
    \frac{\partial}{\partial t}\left(\hat{U}_i+\hat{\Gamma}_1 \hat{A}_{||}\right) &=
    -\left[\hat{\psi}_i,\,\hat{U}_i+\hat{\Gamma}_1 \hat{A}_{||}\right]_{\hat{\perp}} + \\
    + \left[\hat{\Gamma}_1 \hat{A}_{||},\,\frac{1}{2}\,\hat{U}_i^2\right]_{\hat{\perp}}
    +& \tau_i\,\left[\hat{\Gamma}_1 \hat{A}_{||},\,\ln(\hat{N}_i)\right]_{\hat{\perp}}
  \label{eq:4.1_ff4}  
\end{align}
where we denoted that, in the presented case, the gyrocenter density and parallel velocity $\hat{N}_e$ and $\hat{U}_e$ can be approximated to the particle density $n_e$ and velocity $\hat{u}_e$. This approximation holds for electron mass ratios $|\mu_e|=|\frac{m_e}{Z_e m_i}|\ll 1$ on scales of $\rho_s$ and small $\beta$. The main interaction will be between the decaying magnetic field configuration and drift waves, hence this scaling is typically used in nuclear fusion simulations. 
The system of equations is closed by the polarisation equation, presented in the non Oberbeck-Boussinesq arbitrary wavelength form \ref{eq:polff}, and Ampère's law \ref{eq:4.1_ff5}
\begin{align}
    -\frac{1}{\beta}\,\hat{\Delta}_{\hat{\perp}} \hat{A}_{||} = 
    -\hat{n}_e\,\hat{u}_e + \hat{\Gamma}_1\left(\hat{N}_i\,\hat{U}_i\right),
\label{eq:4.1_ff5}
    \\
\hat{\vbm{\nabla}} \cdot [\sqrt{\hat{\Gamma}_0} \{ (\hat{N}_i/\hat{B}^2)\sqrt{\hat{\Gamma}_0} (\hat{\nabla}_{\hat{\perp} }\hat{\phi})\}]=-\sum_zZ_z \hat{\Gamma}_{1,z} \hat{N}_z.\\
\label{eq:polff}
\end{align}

The NOB arbitrary wavelength polarisation equation, can be written as a divergence of the polarisation density $\hat{\textbf{P}}:=\hat{\textbf{P}}_1+\hat{\textbf{P}}_2$ \citep{Held3,Kendl}
\begin{align}
    \sum_z Z_z\hat{N}_z-\hat{\vbm{\nabla}} \cdot \hat{\textbf{P}}=0
\end{align}
with
\begin{align}
    \hat{\textbf{P}}_1&:=-\sum_z  Z_z \hat{\nabla}_{\hat{\perp} }\mu_z \tau_z \hat{N}_z/2,
\\  
    \hat{\textbf{P}}_2&:=-\sum_z  \bigg (\sqrt{\hat{\Gamma}_0} Z_z\mu_z \frac{\hat{N}_z}{\hat{B}^2}\sqrt{\hat{\Gamma}_0} \hat{\nabla}_{\hat{\perp} }\phi \bigg )
\end{align}
where we adopted the second order accurate Padé approximation of \cite{Held2} $\hat{\Gamma}_1\approx \sqrt{\hat{\Gamma}_0}$ and use $ -\hat{\vbm{\nabla}} \cdot \mathbf{P}_1 =\sum_z Z_z (\hat{\Gamma}_1-1)\hat{N}_z.$
The assumptions and simplifications applied to the various models are summed up in table \ref{polapprox}.
The normalized energy balance is
\begin{align}
\label{engtheorem}
\hat{E}_{tot}= \sum_z \int_V dV \bigg ( Z_z\tau_z&\hat{N}_z\ln(\hat{N}_z)+ Z_z\mu_z \hat{N}_z \frac{1}{2}\hat{u}_E^2+\nonumber\\
+Z_z\mu_z\frac{1}{2}&\hat{N}_z \hat{U}_z^2 +\frac{|\hat{\nabla} \hat{A}_\parallel|^2}{2\beta} \bigg )=\\
=\hat{E}_n+\hat{E}_{E\times B}+&\hat{E}_{\hat{\parallel}} + \hat{E}_m + \hat{E}_{\hat{\nu}}\nonumber
\end{align}
and consists of the thermal free energy $\hat{E}_n$, the $E\times B$ energy $\hat{E}_{E\times B}$, the parallel kinetic energy $\hat{E}_{\hat{\parallel}}$ and the magnetic energy  $\hat{E}_m$.
The chosen normalisation aims to capture turbulence on the scales of $\rho_s $, setting the general framework of resolvable scales, which results in only treating the ion FLR-effects. The latter is expected to be relevant only in a situation of large aspect ratio, defined by the ratio of the domain lengths in x- and y-directions, $\hat{L}_{\hat{x}}/\hat{L}_{\hat{x}}$ and turbulent magnetic reconnection \cite{Biancalani}.
\subsection{$\delta$F-limit}
Technically distinguished on the level of the distribution function \cite{ScottV2}, we can recover the \(\delta F\) limit assuming a constant background density $N_0$ and small perturbations $\hat{\tilde{N}}_z$ , substituting $\hat{N}_z=1 + \hat{\tilde{N}}_z$  in the Full-F equations.  We further simplify using $\hat{\tilde{n}}_z \ll \hat{N}_0$, or normalized, $\hat{\tilde{n}}_z\ll1$, ignoring triple non-linearities and inserting the \(\delta F\) polarisation equation
\begin{align}
    \frac{\partial}{\partial t} \hat{\tilde{n}}_e &=
    -\left[\hat{\phi},\,\hat{\tilde{n}}_e\right]_{\hat{\perp}}
    + \left[\hat{A}_{||},\,\,\hat{u}_e\right]_{\hat{\perp}},
    \label{eq:df1}
    \\
    \frac{\partial}{\partial t}  \hat{\tilde{N}}_i &=
    -\left[\hat{\psi},\,\delta \hat{\tilde{N}}_i\right]_{\hat{\perp}} +
    \left[\hat{\Gamma}_1 \hat{A}_{||},\,\,\hat{U}_i\right]_{\hat{\perp}},    \label{eq:df2}
    \\
    \frac{\partial}{\partial t}\left(\hat{u}_e+\frac{1}{\mu_e}\,\hat{A}_{||}\right) &=
    -\left[\hat{\phi},\,\hat{u}_e+\frac{1}{\mu_e}\,\hat{A}_{||}\right]_{\hat{\perp}} - \\
   & -\frac{1}{\mu_e}\,\left[\hat{A}_{||},\,\hat{\tilde{n}}_e\right]_{\hat{\perp}},
  \label{eq:df3}
    \\
    \frac{\partial}{\partial t}\left(\hat{U}_i+\hat{\Gamma}_1 \hat{A}_{||}\right) &=
    -\left[\hat{\psi},\,\hat{U}_i+\hat{\Gamma}_1 \hat{A}_{||}\right]_{\hat{\perp}} + \\
   +& \tau_i\,\left[\hat{\Gamma}_1 \hat{A}_{||},\,\hat{\tilde{N}}_i\right]_{\hat{\perp}},
   \label{eq:df4}
    \\
    -\frac{1}{\beta}\,\hat{\Delta}_{\hat{\perp}} \hat{A}_{||} &=
    -\,\hat{u}_e + \hat{\Gamma}_1\left(\,\hat{U}_i\right),
    \label{eq:df5}
    \\
 \frac{1}{\tau_i}(\hat{\Gamma}_0-1)\hat{\phi} &=  \hat{\tilde{n}}_e - \hat{\Gamma}_1 \hat{\tilde{N}}_i
 \label{eq:df6}
\end{align}
The resulting \(\delta F\) model is in that way similar to an Oberbeck-Boussinesq approximation of the Full-F model. Note that Full-F and Full-F + OB polarisation equation only meet in LWL-approximation \ref{polapprox}. These equations coincide with the model presented by Scott  \cite{ScottV2}. Note, that a major difference between the Full-F and Full-F + OB model are the cubic non-linearities in all evolution equations. 
\begin{table*}[]
    \centering
    \begin{tabular}{|l|l|l|}
    \hline 
    \textbf{GF model}& $\hat{\textbf{P}}_2$ & $\hat{\psi}_2$\\
    \hline 
    Full-F & $-\sum_z\sqrt{\hat{\Gamma}_0} \mu_z Z_z \frac{\hat{N}_z}{\hat{B}_0^2} \sqrt{\hat{\Gamma}_0} \hat{\nabla}_{\hat{\perp}} \hat{\phi} $& $-\mu_z |\hat{\nabla}_{\hat{\perp}} \sqrt{\hat{\Gamma}}_0 \hat{\phi} |^2$\\
    \hline 
    Full-F + OB & $ -\sum_z \mu_z Z_z\frac{\hat{N}_0}{\hat{B}_0^2} \hat{\Gamma}_0 \hat{\nabla}_{\hat{\perp} }\hat{\phi}$& 0\\
    \hline 
        Full-F + LWL & $ - \sum_z \mu_z Z_z \frac{\hat{N}_z}{\hat{B}_0^2} \hat{\nabla}_{\hat{\perp} }\hat{\phi} $& $-\mu_z |\hat{\nabla}_{\hat{\perp} }  \hat{\phi} |^2$\\
        \hline 
  Full-F + OB + LWL & $- \sum_z \mu_z Z_z \frac{\hat{N}_0}{ \hat{B}_0^2} \hat{\nabla}_{\hat{\perp} }\hat{\phi}$& 0\\
    \hline
    \end{tabular}
    \label{polapprox}
    \caption{Long wavelength (LWL) and Oberbeck-Boussinesq (OB) approximation of the polarisation density  $\hat{\textbf{P}}_2$ and polarisation part of the gyrofluid potential $\hat{\psi}_2$ for different models.}
\end{table*}

\section{Numerical implementation and Verification}
\label{num}
\subsection{Polarisation equation and Ampère's law}
\label{fieldeq}
Since GREENY was developed with TIFF as a foundation, several solvers are used in the same way as implemented in TIFF \cite{Kendl}. The main solvers to mention are the Teague's Method and the Preconditioned-Conjugate-Gradient-Method (PCG), both used to invert the polarisation equation \citep{Teague, Fisicaro}. For details we refer to TIFF, but the implementation was tested separately to ensure its correctness. All model versions of the polarisation equation, shown in table \ref{polapprox}, are implemented for any choice of solver.

Besides the polarisation equation, the inversion of Ampère's law is of special interest. Starting with the canonical momenta $\hat{A}_{\hat{\parallel},i}^*=\hat{U}_i+\hat{\Gamma}_1 \hat{A}_{\hat{\parallel}} $ and $\hat{A}_{\hat{\parallel},e}^*=\hat{u}_e+\frac{1}{\mu_e}\hat{A}_{\hat{\parallel}}$
which now are substituted for $\hat{U}_z$ in Ampère's law

\begin{align}
-\frac{1}{\beta}\hat{\Delta}_{\hat{\perp}} \hat{A}_{\hat{\parallel}}=-\hat{n}_e(\hat{A}_{\hat{\parallel},e}^*-\frac{1}{\mu_e}\hat{A}_{\hat{\parallel}})+\hat{\Gamma}_1 (\hat{N}_i(\hat{A}_{\hat{\parallel},i}^*-\hat{\Gamma}_1 \hat{A}_{\hat{\parallel}})) 
\end{align}
Depending on the solver of choice one now has to decide how to treat the equation. A spectral solver would need a deconvolution algorithm for the Full-F quantities as the equation reads
\begin{align}
\frac{1}{\beta}k^2 \hat{A}_{{\hat{\parallel}},\hat{k}} +\frac{1}{\mu_e}\hat{n}_e \ast \hat{A}_{{\hat{\parallel}}\hat{k}} &-  \frac{\hat{N}_i }{1+k^2/2} \ast  \bigg(\frac{1}{1+k^2/2} \hat{A}_{{\hat{\parallel}}\hat{k}} \bigg )  =\\
=\hat{n}_e \hat{A}_{\hat{\parallel},e}^*&-\Gamma_1(\hat{N}_i \hat{A}_{\hat{\parallel},i}^*)=RHS
\end{align}
and we used the Padé form of the Gyro-averaging operator and $\ast$ denotes the convolution.  In the \(\delta F\) limit, the spectral inversion is straight forward as the equation simplifies to $\bigg (\frac{1}{\beta}\hat{\Delta}_{\hat{\perp}}+\frac{1}{\mu_e} -  \hat{\Gamma}_1^2 \bigg ) \hat{A}_{{\hat{\parallel}}\hat{k}}= \hat{A}_{\hat{\parallel},e}^*-\Gamma_1( \hat{A}_{\hat{\parallel},i}^*)$, where no deconvolution is needed. A straight forward spectral inversion of the Full-F equation, without deconvolution, therefore only holds in a situation of fluctuations in density $\hat{N}_z/\langle \hat{N}_z\rangle_{\hat{V}}\ll 1$, where $\langle \cdot \rangle_{\hat{V}}=\frac{1}{\hat{V}}\int_{\hat{V}}d\hat{V}(\cdot)$ is the volume average. This works rather well in many cases, but as a nature of the Full-F model we are interested in situations where the mean deviation of the densities is as large as the initial or equilibrium configuration. Nevertheless, the here discussed test cases displayed rather small density fluctuations, and hence GREENY treats the Full-F RHS with constant densities . To avoid this problem, a SOR (successive over-relaxation) scheme was implemented to invert the Full-F Ampère's law.  \\

While the solvers for the polarisation routine are, in one or the other way, available for all options (e.f. Full-F/\(\delta F\), arbitrary wavelength/lwl), the routine for solving Ampère's law has to be chosen to match the initial condition. In algorithm \ref{alg:amp} we show the treatment of canonical momentum to obtain the parallel vector potential $\hat{A}_{\hat{\parallel}}$. In this pseudo-code the variables $isolver$ and $ifullf$ are used, to create the different solver configurations. $\hat{A}_{{\hat{\parallel}},z}^*=\hat{U}_z+\frac{1}{\mu_z}\hat{A}_{{\hat{\parallel}},z}$ is the canonical momentum, obtained from the RHS of eq. \ref{eq:4.1_ff3} and \ref{eq:4.1_ff4} and $f(\cdot)$ denotes the LHS or the product of $\hat{A}_{\hat{\parallel}}$ with field-variables, as a function of the arguments. For spectral methods, a suitable Fourier coefficient $c_{ff}$ is chosen  \\
\begin{figure}[ht]
\centering
\begin{minipage}{.7\linewidth}
\begin{algorithm}[H]
\KwData{$\hat{A}_{{\hat{\parallel}},i}^*,\ \hat{A}_{{\hat{\parallel}},e}^*,\ \hat{n}_e,\ \hat{N}_i,\ \hat{u}_e, \ \hat{U}_i, \ \hat{A}_{\hat{\parallel}}$}
\uIf{$isolver < 0 || isolver = 3 || isolver = 5$}
{
	\uIf{$ifullf = 0$}
	{
	RHS$=f(\hat{A}_{{\hat{\parallel}},e}^*,\hat{A}_{{\hat{\parallel}},i}^*,\hat{n}_e, \hat{N}_i)$\\
	INS$=f(\hat{n}_e, \hat{N}_i)$
	}\uElseIf{$ifullf = -1$}
	{
	RHS$=f(\hat{A}_{{\hat{\parallel}},e}^*,\hat{A}_{{\hat{\parallel}},i}^*,\hat{n}_e, \hat{N}_i)$\\
	INS$=f(\hat{n}_e=1, \hat{N}_i=1)$
	}
	\textit{successive over-relaxation method }\\
  \Return $\hat{A}_{\hat{\parallel}}$\;
}
\Else
{

\uIf{$ifullf = 0$}{
	\uIf{$\tau_i=1$}{$c_{ff}(\hat{N}_0,\hat{\Gamma}_{0,i})$ }
	\Else{ $c_{ff}(\hat{N}_0)$}
	RHS$=f(\hat{A}_{{\hat{\parallel}},e}^*,\hat{A}_{{\hat{\parallel}},i}^*,\hat{n}_e, \hat{N}_i)$\\
	INS$=f(\hat{n}_e=1, \hat{N}_i=1)$\\
	\textit{fftw}(RHS,INS,$c_{ff}$)
	}\uElseIf{$ifullf = -1$}
	{
	RHS$=f(\hat{A}_{{\hat{\parallel}},e}^*,\hat{A}_{{\hat{\parallel}},i}^*)$\\
	INS$=f(\hat{n}_e=1, \hat{N}_i=1)$\\
	\textit{fftw}(RHS,INS,$c_{df}$)
	}
  \Return $\hat{A}_{\hat{\parallel}}$\;
}
\caption[Ampère]{Procedure of the selection of the inversion algorithm for Ampère's law.}\label{alg:amp}
\end{algorithm}
\end{minipage}
\end{figure}
The major challenge in algorithm \ref{alg:amp} lies in the Full-F treatment of Ampère's law. The SOR-solver can be used for arbitrary amplitudes in the densities (Full-F) but ignores the $\hat{\Gamma}_1$ in both,  $\hat{\Gamma}_1(\hat{N}_i\hat{\Gamma}_1)$ and $\hat{\Gamma}_1(\hat{N}_i \hat{A}_{{\hat{\parallel}},i}^*)$. The spectral solver on the other hand is only valid for perturbations in the densities $\hat{\tilde{n}}/\hat{N}_0\ll 1$ (close to the \(\delta F\) approximation) but all FLR-terms are included. The selection of the solver is hence driven by the problem at hand since some Full-F simulations, although defined by arbitrary density fluctuations, might still obey this limit (e.g. Harris-Sheet magnetic reconnection).

\begin{center}
\begin{figure}[htp]
\centering
  \includegraphics[scale=0.65]{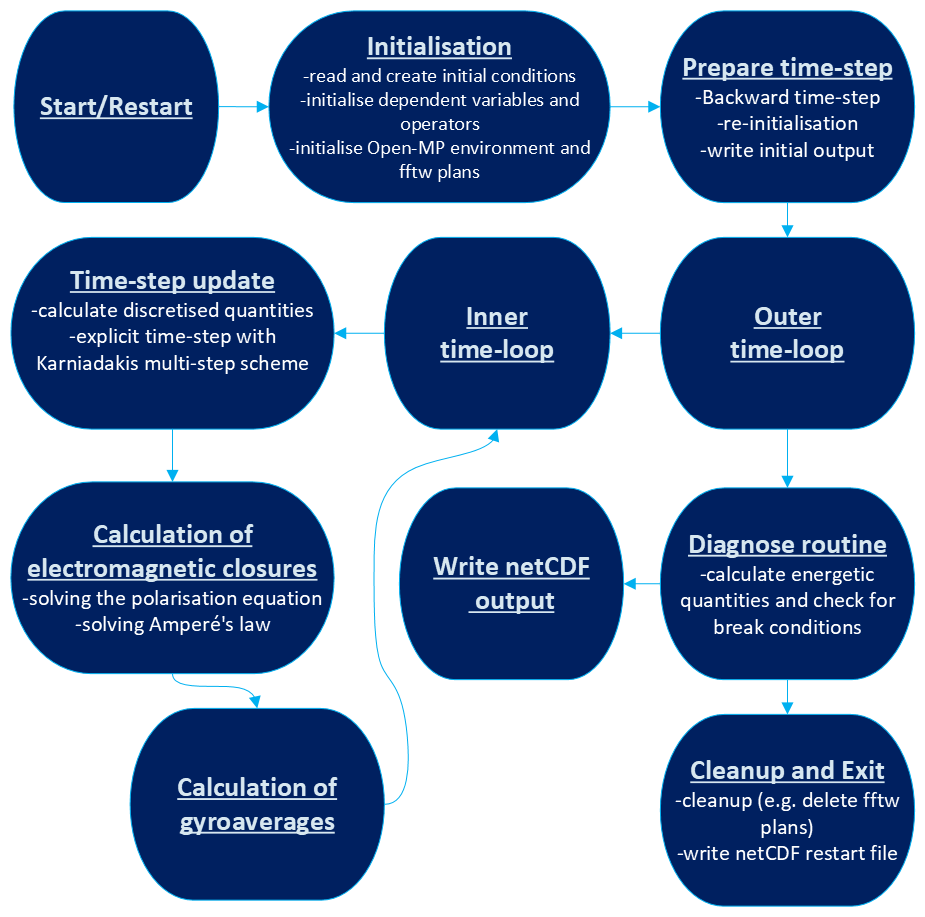}
  \caption{Flow chart of the simulation algorithm and relevant subroutines.}
  \label{Flow}
\end{figure}
\end{center}

Figure \ref{Flow} shows the flow chart of the simulation, the initialization routine and the division in an "inner" and an "outer" time-loop. The inner time-loop first solves the time step with a Karniadakis multistep scheme \cite{Karniadakis} and then proceeds to calculate the electromagnetic potentials and gyroaveraged quantities.  To avoid a delayed onset of the reconnection we backwards integrate the initial conditions in time with an Euler-Scheme to obtain the correct previous time-step states, needed for the multistep scheme. The outer time-loop runs a diagnose routine, calculating energies, evaluating break conditions and creating the netCDF output.\\

\begin{center}
\begin{figure}[h]
\centering
  \includegraphics{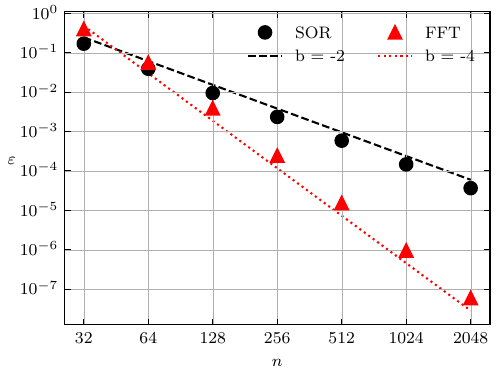}
  \caption{Maximum error scaling for the inversion of Amperé's law for SOR (discs) $\epsilon_{SOR} \propto h^2$ and FFT (triangles) $\epsilon_{SPEC} \propto h^4$. }
  \label{scalingAmp}
\end{figure}
\end{center}

 For the evaluation of gradient or Laplacian operators, both spectral (SPEC) methods and finite difference schemes (FDS) are available, while Poisson brackets are evaluated with the Arakawa scheme \cite{Arakawa}. The solvers have been tested with analytical solutions, where suitable functions were chosen to mimic a realistic inversion process. 
To invert Amperè's  law (Eq. \ref{eq:4.1_ff5}) for $\hat{A}_{\hat{\parallel}}$ one can choose between Successive Over-Relaxation (SOR) or a Fast Fourier Transform-Method (FFTW) \cite{FFTW}, although the latter is only accurate in the \(\delta F\) limit, avoiding spectral convolution. Figure  \ref{scalingAmp}  shows the scaling of the maximum error $\epsilon$ with grid resolution, where $n_{x,y}$ denotes the number of grid points. Note that for SPEC and periodic boundary conditions, the RHS of eq.~\ref{eq:4.1_ff5} is used on a doubled domain (mirrored inversely) to avoid jumps and kinks. 
 For the given Full-F problem of the induction equation, $(\hat{\Delta}_{\hat{\perp}} + f(\vbm{\hat{x}}))a(\vbm{\hat{x}})=g(\vbm{\hat{x}})$ we chose the test functions to be
\begin{align}
f(\vbm{\hat{x}})=-(4-3\sin(\hat{x}^2)),\\
g(\vbm{\hat{x}})=(4 \hat{x}^2-4)\exp(-\hat{x}^2)+f(\vbm{\hat{x}})\exp(-\hat{x}^2),\\
a_{analytic}(\hat{\vbm{x}})=\exp(-\hat{x}^2)
\end{align}
where $|\vbm{\hat{x}}|=\sqrt{\hat{x}^2+\hat{y}^2}$. One can vary $f(\vbm{x})$ and, in the easiest case, use a constant, but it must strictly be negative to avoid spectral singularities. The spectral solvers were tested with $f(\vbm{\hat{x}})=const.$, which only holds for small variations of $f(\vbm{\hat{x}})$, referring to small variations in $\hat{n}_e/\hat{N}_i$ and hence only holds in the \(\delta F\) limit. However one can implement a deconvolution algorithm to avoid this simplification. The local error $\tilde{\epsilon}(\vbm{\hat{x}})$ was obtained by comparing the analytical solution $a_{analytic}$ with the calculated $a(\vbm{\hat{x}})$ locally.
To obtain a scaling parameter a function $\epsilon=a+c\exp({d})$ was fitted  for the maximum error of the solution $f(\vbm{\hat{x}})$,  $\epsilon_\infty=\lVert f_{num}-f_{analytic} \rVert_\infty$ and the result, compared to $\exp({b})$,  is presented in a log-lin plot \ref{scalingAmp}. As expected from the discretization involved in the corresponding method, the SOR solver is approximately of step size order $h^2$ while the FFT scales with $h^4$.\\

For the problem stated by the polarisation equation 

\begin{align}
\vbm{\hat{\nabla}} \cdot (f(\vbm{\hat{x}})\hat{\nabla}_{\hat{\perp}} a(\vbm{\hat{x}}))=g(\vbm{\hat{x}})
\end{align} 

we chose the test functions to be

\begin{gather}
f(\vbm{\hat{x}})=1+a_0 \sin(\hat{x} m \hat{k}_{\hat{x}})\sin(\hat{y} m \hat{k}_{\hat{y}}),\\
g(\vbm{\hat{x}})=f(\partial^2_{\hat{x}} a_{\mathrm{ana}}+\partial^2_{\hat{y}} a_{\mathrm{ana}})+\partial_{\hat{x}} f\partial_{\hat{x}} a_{\mathrm{ana}}+\\
 +\partial_{\hat{y}} f\partial_{\hat{y}} a_{\mathrm{ana}},\\
a_{\mathrm{ana}}(\vbm{\hat{x}})=a_1\sin(\hat{x} m\hat{k}_{\hat{x}})\sin(\hat{y} m\hat{k}_{\hat{y}}).
\end{gather}

\begin{center}
\begin{figure}[h]
\centering
  \includegraphics[trim=0 0 0 0,clip,scale=1]{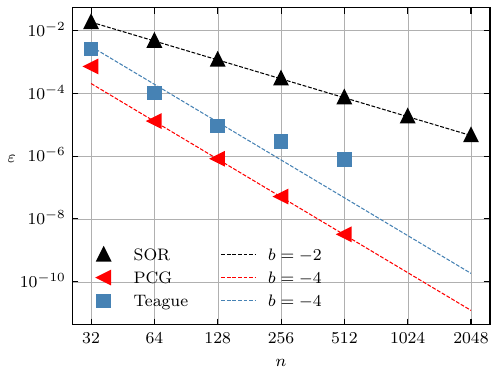}
  \caption{Maximum error scaling of the inversion of the polarisation equation for SOR (upright triangle) $\epsilon_{SOR} \propto h^2$, spectral (iterative Teague)\cite{Kendl} method (squares) $\epsilon_{SPEC} \propto h^4$ and the PCG (tilted triangle) $\epsilon_{PCG} \propto h^4$. }
  \label{ErrorOrderingPaper}
\end{figure}
\end{center}

 As shown in figure \ref{ErrorOrderingPaper}, both the spectral (iterative Teague with 20 iterations) and the iterative PCG solver's error scaled with $\epsilon_{PCG} \propto h^4$ while the SOR again scaled with $\epsilon_{SOR} \propto h^2$.

\subsection{Conserved quantities}
\begin{center}
\begin{figure}[h]
\centering
  \includegraphics[trim=1 1 1 1,clip]{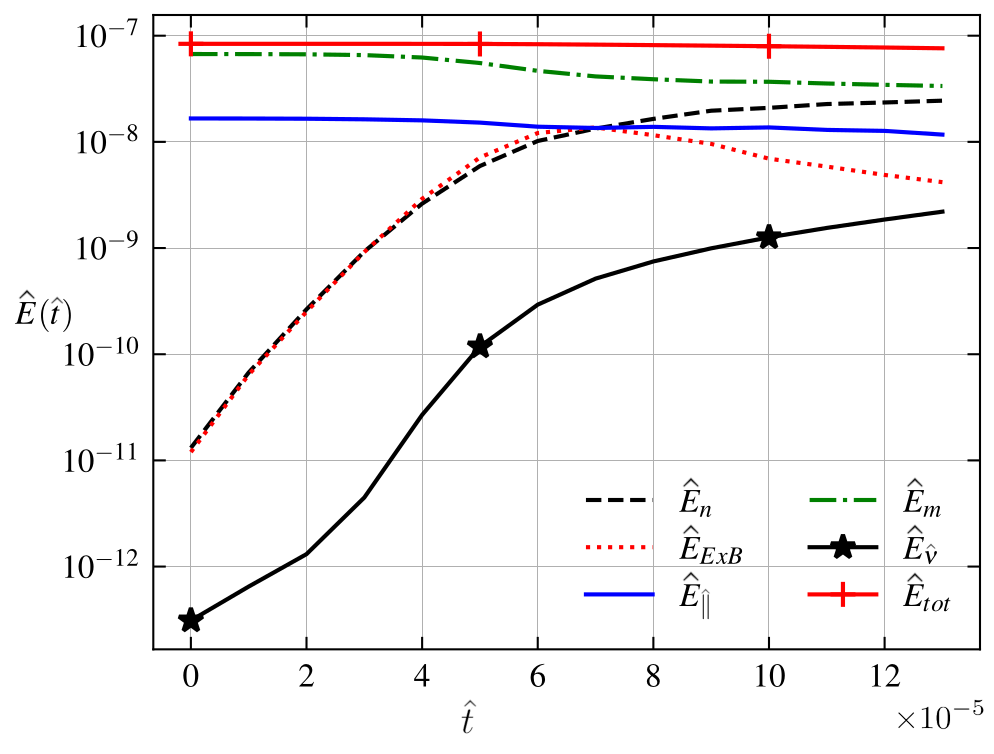}
  \caption{Time evolution of energies for a Full-F Harris-sheet reconnection. The magnetic energy $E_m$ is converted into kinetic $E_{E \times B}$ and thermal free $E_n$ energy. At high enough resolution ($n_x \times n_y=1024 \times 1024$) the hyperviscous sink $E_{\hat{\nu}}$ remains low enough to not interfere drastically with the reconnection process.}
  \label{EnergyCons2}
\end{figure}
\end{center}

The normalized system energy evolution for a Harris sheet initial condition is shown in fig. \ref{EnergyCons2}. As the magnetic energy $\hat{E}_m$ decays in the reconnection event, the perpendicular kinetic  $\hat{E}_{E\times B}$ and thermal free energy $\hat{E}_n$ component rises. As a necessary sink, this energy, and the parallel kinetic energy $\hat{E}_{\hat{\parallel}}$, dissipate via the small scale hyperviscosity $\hat{E}_{\hat{\nu}}$ (subgrid dissipation). This dissipative drive is time integrated and part of the total energy balance $\hat{E}_{tot}$ \ref{engtheorem}.
\begin{align}
\frac{d \hat{E}_{\hat{\nu}}}{d \hat{t}}&=\sum_z \int_{\hat{V}} d\hat{V} Z_z\{[\tau_z(1+\ln(\hat{N}_z))+Z_z\hat{\psi}+\frac{1}{2}\mu_z \hat{U}_z^2]
\nonumber
\\ &\cdot
(-\hat{\nu} \hat{\Delta}^2_{\hat{\perp}} \hat{N}_z)
+\mu_z \hat{U}_z \hat{N}_z\hat{\Lambda}_\xi
\}.
\end{align}

Note that this subgrid dissipation introduces a diffusion, setting the CFL limit for the simulation. In the absence of ohmic resistivity, this diffusion, in combination with the electron inertia, drives the magnetic reconnection.  To obtain good numerical results, one has to investigate the influence of the hyperviscosity in a resolution study. Its scaling parameter $\hat{\nu}$ should be chosen large enough to maintain numerical stability and prevent granular morphology, but small enough, to not distort the physics. As we focus on non-driven systems in this work, the purpose of the subgrid dissipation is of secondary importance and the corresponding hyperviscosity factor $\hat{\nu}$
\begin{align}
\hat{\Lambda}_\xi=-\hat{\nu} \hat{\Delta}^2_{\hat{\perp}}(\xi)
\end{align}
for a field $\xi$.
can be set relatively small ($\hat{\nu}<10^{-7}$).  Its effect on the initial phase of the reconnection should be vanishing due to the $\hat{k}^4$ scaling. However, it will play a role in the pinched current sheet situation, when the width $\hat{L}_{\hat{\perp}}$ becomes relevantly narrow and hence $\frac{\rho_s}{\hat{L}_{\hat{\perp}}}=1$. As this is closely related to the resolution, we will test the grid convergence together with the hyperviscosity in section \ref{examples} on Harris-sheet magnetic reconnection.

In the next step we want to investigate the total energy of the system (eq. \ref{engtheorem}) and its time conservation. For this purpose we used the Harris-Sheet (eq. \ref{harrinit}) with a box stability parameter of $\Delta '=9.79$, resolutions of $\frac{\hat{L}_{\hat{x}}}{n_y}=\frac{\hat{L}_{\hat{x}}}{n_x}=  \frac{1}{32}$  and a hyperviscosity parameter of $ \hat{\nu}= \{ 10^{-6}, 10^{-9} \}$. 
 Figure \ref{EnergyCons2} shows the time evolution of the energetic quantities. The magnetic energy decreases in the reconnection event and the energy is converted into thermal free energy and perpendicular kinetic energy $\hat{E}_{E \times B}$. The resulting small scale turbulence is then dissipated via the hyperviscous drive.
We want to minimize the effect of the hyperviscous dissipation on the reconnection rate, but in cases with small scale turbulent structures, the absence of proper dissipation will lead to unphysical noise and hence again a disturbance of the reconnection process. The choice of $\hat{\nu}$ has to be assessed for each simulation, but we suggest to take it as small as possible, suppressing structures that accumulate on the grid level. Nevertheless, similar models have been shown to display a more complex interplay between hyperviscosity and the physics, hence a more detailed study is left out for future work. \citep{Grander} \\

\begin{center}
\begin{figure}[h]
\centering
  \includegraphics[trim=0 0 0 0,clip,scale=0.5]{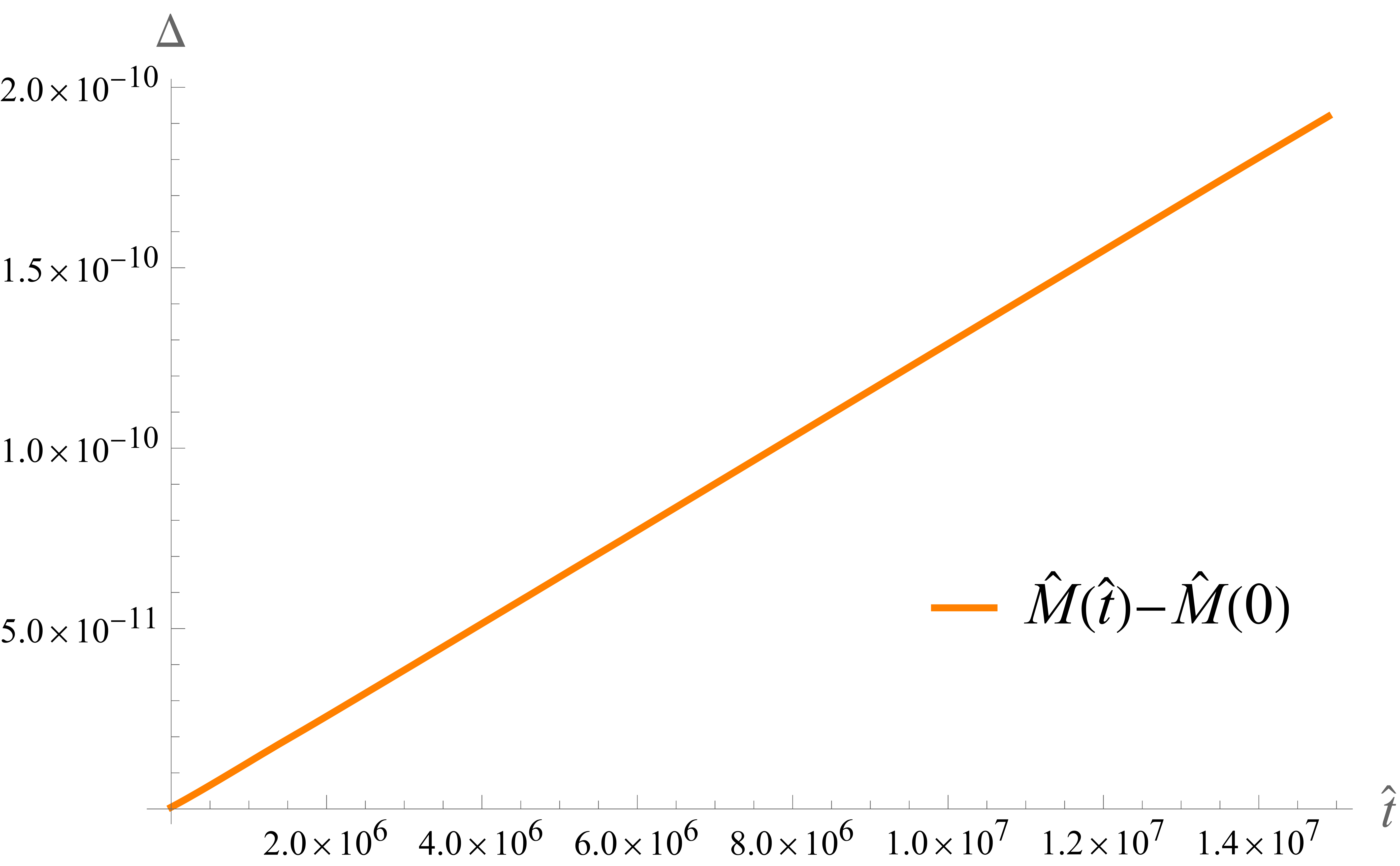}
    \includegraphics[trim=0 0 0 0,clip,scale=0.5]{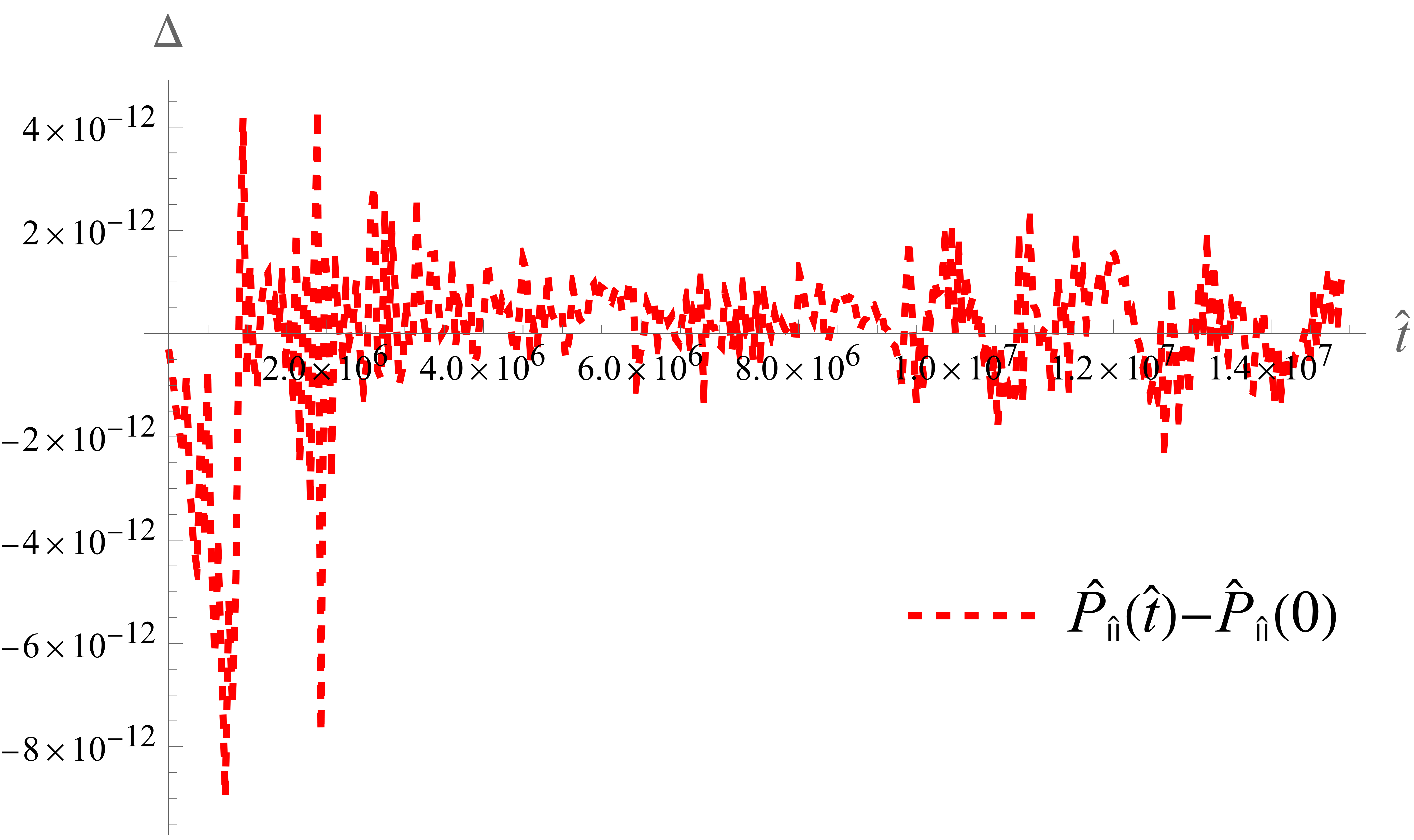}
  \caption{Change $\hat{P}_{\hat{\parallel}}(\hat{t})-\hat{P}_{\hat{\parallel}}(\hat{t}=0)$ of total parallel momentum and $\hat{M}(\hat{t})-\hat{M}(\hat{t}=0)$ for particle number $\hat{M}$. }
  \label{cons}
\end{figure}
\end{center}

As shown in figure \ref{cons} the system particle number \cite{Scott,Scott1}

\begin{align}
\hat{M}=\sum_z\int_{\hat{V}} d\hat{V} \hat{N}_z
\end{align}
and the total parallel canonical momentum

\begin{align}
\hat{P}_{\hat{\parallel}}=\sum_z\int_{\hat{V}} d\hat{V}\bigg ( \mu_z \hat{N}_z \hat{U}_{z}+\hat{A}_{{\hat{\parallel}},z} \bigg )
\end{align}

are both conserved to a reasonable degree. To verify the code's applicability to physical problems within the typical fluid framework we first investigated the fundamental vortex interaction as the foundation of plasma turbulence behaviour. Fundamental vortex interaction in an isothermal low $\beta$ gyrofluid, including FLR effects, was studied extensively by Kendl \cite{KendlVort}.  To prove the correct implementation of the $\tau_i \neq 0$ framework, we were able to reproduce these fundamental observations. This is of special importance, as in the electron-induced current sheet with small $\Delta'_{box}=2 (1/\hat{k}_{\hat{y}}-\hat{k}_{\hat{y}})$, the ion FLR effects are expected to be negligible \cite{Biancalani}. 

\subsection{Granalysis}
Granalysis is a script written in python that can be used for conveniently plotting and animating the output data of GREENY. Furthermore, it can be used to calculate reconnection rates, the energy statistics, as well as more specific use cases that can be viewed by calling \texttt{/path/to/granalysis -h}.\\
For example, in order to create an animation from an output file of GREENY, one can call granalysis like \texttt{/path/to/granalysis . -o . -a Ap}. The first argument tells granalysis where greeny's output file is, the flag \texttt{-o} where the animation will be saved (in this case as an .mp4 file) and lastly \texttt{-a} calls the animation function of granalysis on the variable \textit{Ap}.
\section{Inital  conditions}
\label{inibound}
Most system parameters of interest are set via the input file, including environmental parameters such as \texttt{numthreads}, which could also be set in terminal. A detailed explanation of the input file and parameters can be found in the git documentation.

The following paragraph lists the most relevant and unique initial conditions and explains what has to be considered when constructing the input file. Each initial condition is uniquely numbered with $\texttt{incon}\in [1,1000]$ without any distinct scheme or order.
\begin{enumerate}
\item[•] \texttt{incon}=(10 \textrm{\&} 11), \texttt{"Harris-Sheet magnetic reconnection"}: Initializes a central current sheet and the corresponding magnetic field. When \texttt{incon}=10 is chosen, the outermost $\cos$ is omitted.
\begin{gather}\label{harrinit}
\hat{A}_{\hat{\parallel}} =\beta  \bigg[\frac{A_0}{\mathrm{cosh}(\sigma  4 \pi  \hat{x}/\hat{L}_{\hat{x}})^2}+ A_1  \cos(m_{\hat{y}}  2 \pi \hat{y}/\hat{L}_{\hat{y}})  \bigg ] \nonumber \\ \cdot \cos(\pi \hat{x}/\hat{L}_{\hat{x}}),\\
\hat{u}_e=\frac{1}{\hat{N}_0\beta}\hat{\Delta}_{\hat{\perp}} \hat{A}_{\hat{\parallel}} 
\end{gather}
\texttt{parameters}:\{Amplitude of equilibrium sheet $:=A_0$, perturbation amplitude$:=A_1$, sheet width $:=\sigma$, perturbation mode number $:=m_y$\}

\item[•] \texttt{incon}=13, \texttt{"Magnetic Islands"}: Initializes a vector potential with multiple magnetic islands, analogous to Stanier et al. \cite{Stanier}.
\begin{gather}
\hat{A}_{\hat{\parallel}}= \beta  \bigg \{  A_0 \frac{\hat{L}_{\hat{x}}}{4 \pi}\ln  \bigg [ \mathrm{cosh} \bigg ( \sigma \frac{2 \pi \hat{x}}{\hat{L}_{\hat{x}}} \bigg )  +
 \epsilon  \cos \bigg (\eta \frac{2 \pi \hat{y}}{\hat{L}_{\hat{x}}} \bigg ) \bigg ]+
\nonumber\\
+ A_1  \cos \bigg ( \frac{2 \pi m_y \hat{y}}{\hat{L}_{\hat{x}}} \bigg) \bigg \} \cos(\pi \hat{x}/\hat{L}_{\hat{x}}), \\
\hat{u}_e=\frac{1}{\hat{N}_0 \beta}\hat{\Delta}_{\hat{\perp}} \hat{A}_{\hat{\parallel}}
\end{gather}
\texttt{parameters}:\{Amplitude of equilibrium islands $:=A_0$, perturbation amplitude$:=A_1$, island width $:=\sigma$, island distance $:=\eta$, perturbation mode number $:=m_y$\}

\item[•] \texttt{incon}=18, \texttt{"Magnetic Vortex interaction"}: Initializes two magnetic Gaussian vortices and calculates a consistent electric potential $\phi$ and ion gyrocenter density $N_i$. Due to the inversion process needed for the calculation of $\phi$ and $N_i$, this initialization is limited to \(\delta F\) and comparable simulations.
\begin{gather}
\hat{n}_e=A_3 \exp \left [-\frac{(\hat{x}-\hat{x}_0)^2}{(\sigma  \hat{L}_{\hat{x}})} \right ] \exp \left [-\frac{(\hat{y}-\hat{y}_0)^2}{(\sigma  \hat{L}_{\hat{x}} )} \right] , \\
\hat{A}_{\hat{\parallel}}=\beta  A_0  \exp \left [-\frac{(\hat{x}-\hat{x}_0)^2}{(\sigma  \hat{L}_{\hat{x}} )} \right ] \exp \left [-\frac{(\hat{y}-\hat{y}_0)^2}{(\sigma  \hat{L}_{\hat{x}})} \right]  , \\
\hat{u}_e=\frac{1}{\hat{N}_0  \beta}\hat{\Delta} _{\hat{\perp}} \hat{A}_{\hat{\parallel}}
\end{gather}
\texttt{parameters}:\{Amplitude of magnetic blob $:=A_0$, vortex density-amplitude $:=A_3$, blob width $:=\sigma$, blob location $:=\hat{x}_0$ and $\hat{y}_0$\}

\item[•]  \texttt{incon}=2, \texttt{"Restart initial condition"}: Restarts from folder named "\texttt{greeny\_out\_\#}" which is either the highest number \# in the local directory (\texttt{restarttarget} $<0$) or a selected number \# (\texttt{restarttarget} $>0$).
\texttt{parameters}:\{•\}
\end{enumerate}

\section{Computational examples}
\label{examples}
\subsection{Harris-sheet magnetic reconnection}
\begin{figure}[htp]
\centering
  \includegraphics[trim=2 0 5 0,clip,width=0.55\columnwidth]{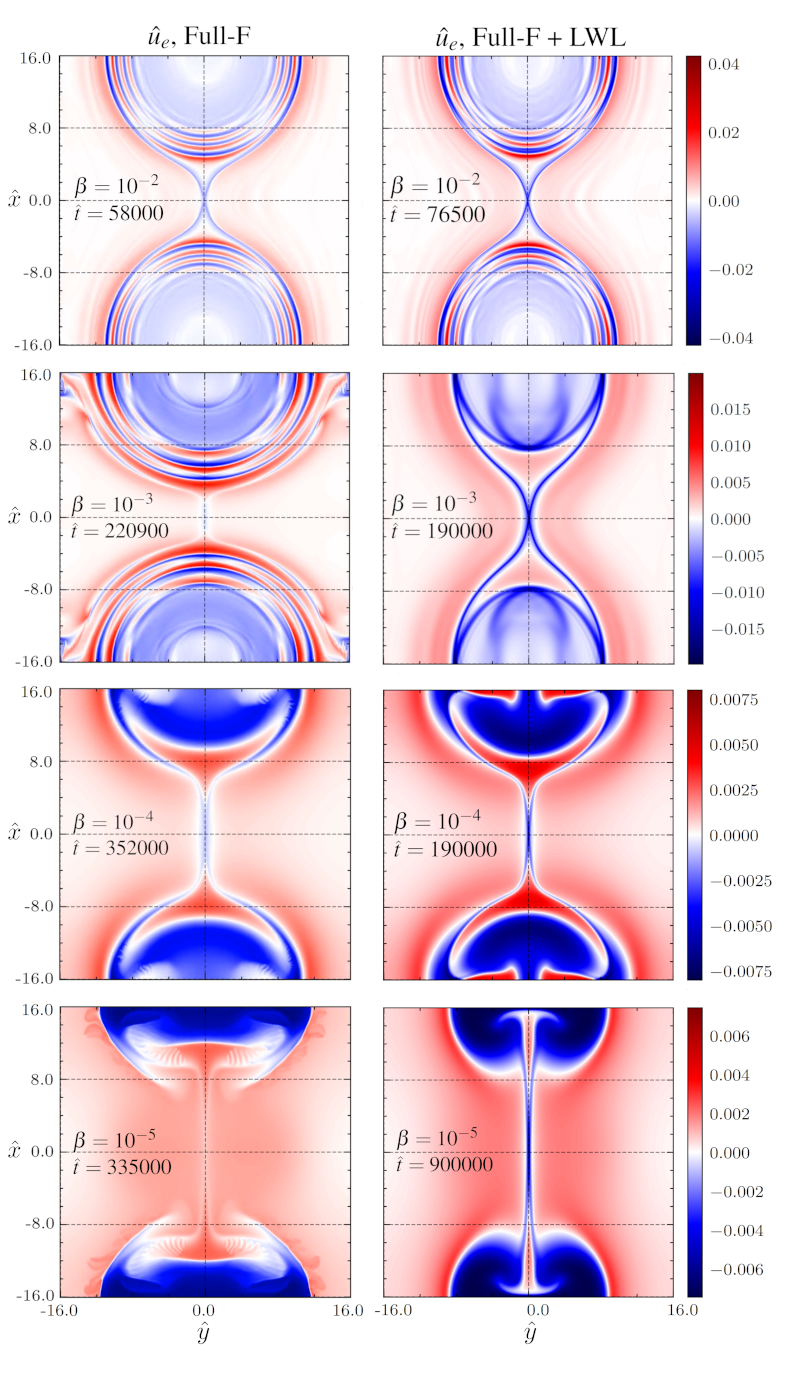}
  \caption{Electron parallel velocity $\hat{u}_e(\vbm{\hat{x}})$ for $\beta \in \{10^{-5},10^{-4},10^{-3},10^{-2} \}$, for the Full-F long wavelength limit (right) and arbitrary wavelength polarisation (left). The elongation of the reconnection zone  increases with decreasing $\beta$.}
  \label{betashape2}
\end{figure}

\begin{figure}[htp]
\centering
    \includegraphics[trim=4 0 2 0,clip,width=0.5\columnwidth]{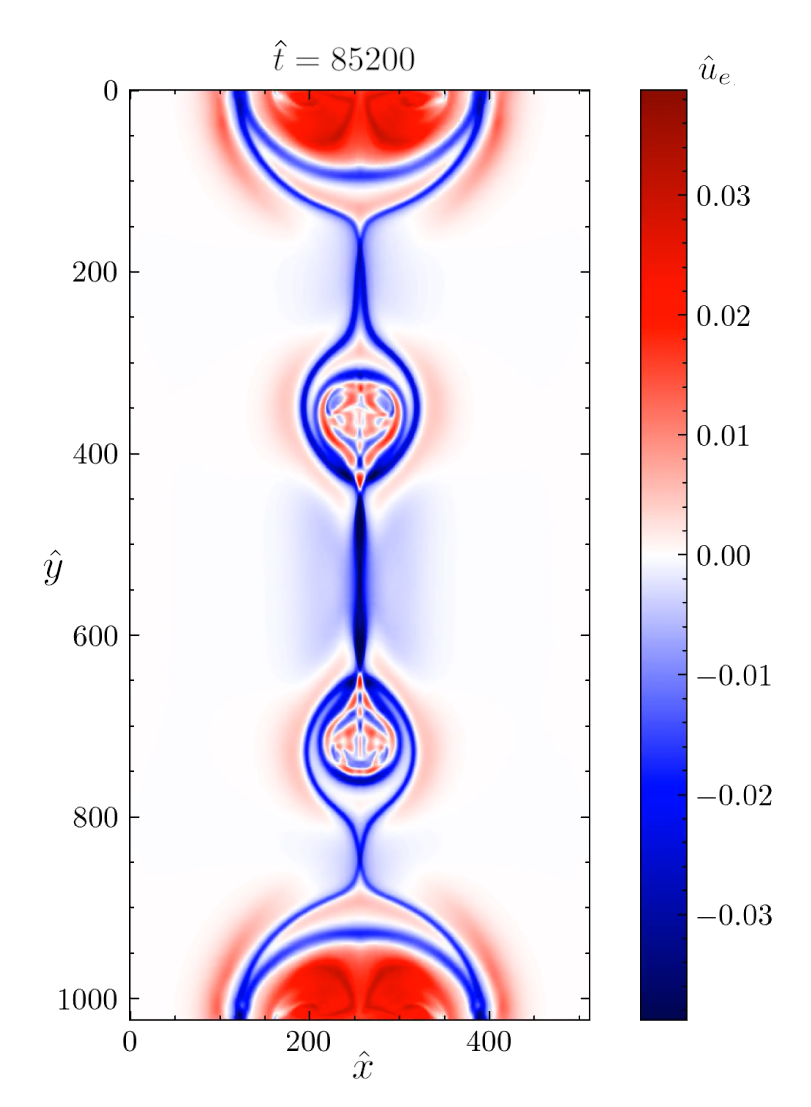}    
  \caption{Parallel electron velocity $\hat{u}_e (\hat{\vbm{x}})$ for an ion-current induced magnetic reconnetion event with aspect ratio $\frac{\hat{L}_{\hat{y}}}{\hat{L}_{\hat{x}}}=2$ and $\tau_i=1$. The electron current shows two plasmoids growing, due to the ion FLR-effects.}
  \label{plasmo}
\end{figure}

 In figure \ref{betashape2}, one can see the shape of the reconnection zone varying with $\beta$. It is noticeable, that the reconnection zone is relatively X-shaped for higher values of $\beta$, while the reconnection zone is clearly elongated (Y-shaped) for lower values, which matches with previous observations \cite{Biancalani}. In contrast to the LWL-approximated case (left), the non OB reconnection displays fine structures on length scales of $\rho_s$ and resolving them adds to the already high demand on resolution. A first study of fundamental magnetic reconnection processes was done using GREENY by A. Stürz during his master thesis \cite{Stürz}. Figure \ref{plasmo} shows an ion current $\hat{U}_i\hat{N}_i$ induced reconnection with an aspect ratio of $\frac{\hat{L}_{\hat{y}}}{\hat{L}_{\hat{x}}}=2$ and warm ions $\tau_i=1$. Due to ion-FLR effects, the difference in ion and electron potentials $\hat{\psi}_i$ and $\hat{\psi}_e$ decouple the motion on small scales and lead to the formation of two plasmoids.

\begin{figure}[htp]
\centering
  \includegraphics[trim=1 1 1 1,clip,width=0.5\columnwidth]{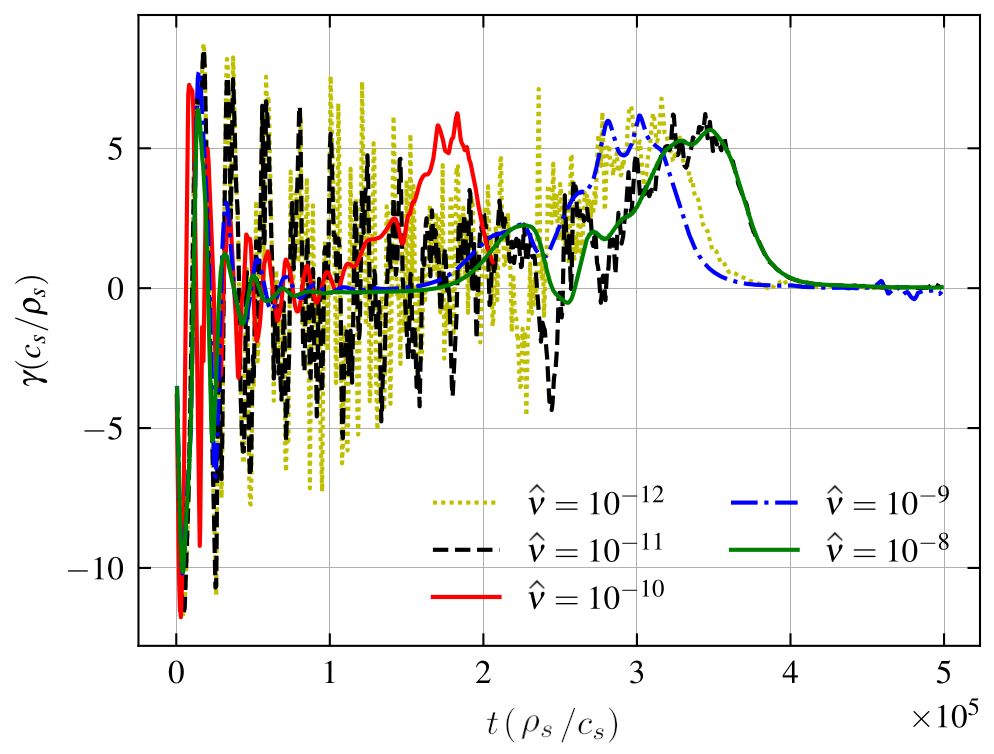}
  \caption{Convergence of reconnection rates $\gamma$ for the Full-F Harris-sheet at $\beta=5\cdot 10^{-3}$ and $\Delta_{box}'=0.15$ with changing hyperviscous scaling parameter $\hat{\nu}$. Although some overlap for the main reconnection event, the peak after the transient phase, was observed. The onset of the event varies, without any clear connection to $\hat{\nu}$, but it becomes clear, that subgrid dissipation is highly influential on the reconnection event.}
  \label{convnu}
\end{figure}

 To check the influence of the hyperviscous scaling parameter $\hat{\nu}$ we scanned the magnetic reconnection rate 

\begin{align}
\gamma=\frac{d}{dt}\ln(\delta \hat{A}_{\hat{\parallel}})=\frac{d}{dt}  \ln|\hat{A}_{\hat{\parallel}}(\vec{x}_0,t)-\hat{A}_{{\hat{\parallel}},eq}(\vec{x}_0)| 
\end{align} 
over a range of $\hat{\nu} =10^{-12}$ up to $10^{-6}$, shown in figure \ref{convnu}. In the Full-F cases with $\beta=5\cdot 10^{-3}$ and $\Delta_{box}'=0.15$, the system goes through a relatively long transient phase, characterized by island formation and merging. Although an increasing $\hat{\nu}$ is shown to shorten this intermittent phase, no direct influence to neither the general shape, height and onset of the main reconnection event, defined by the highest peak following the transient phase, is observed. We further observe that for $\hat{\nu}>10^{-6}$ the situation becomes linearly stable, as the initial perturbation is flattened out completely, and both simulations remained stable on the relevant time scale. During the transient phase, a change from $m=4.9$ to something close to $m=1$ is observed, but the exact location of the X-point varies, and again no connection to $\hat{\nu}$ was visible. From $\hat{\nu}=10^{-11}$ downwards, the morphology of $u_e$ and $n_e$ becomes more and more granular. This lack of subgrid dissipation might easily influence the ongoing reconnection, and hence a larger $\hat{\nu} \in \{ 10^{-7},10^{-9}\}$ is reasonable. We highly recommend testing each case for convergence on an individual basis and on multiple levels, e.g. check granularity of morphology and compare to ratio of hyperviscous drive and total energy.

One should keep in mind that convergence might be more complex after all, since the conservation of magnetic field lines is fundamentally broken due to the discretization \cite{Morillo}.

\subsection{Magnetic islands and vortices}
\begin{figure}[htp]
\centering
  \includegraphics[trim=3 2 2 0,clip,width=0.5\columnwidth]{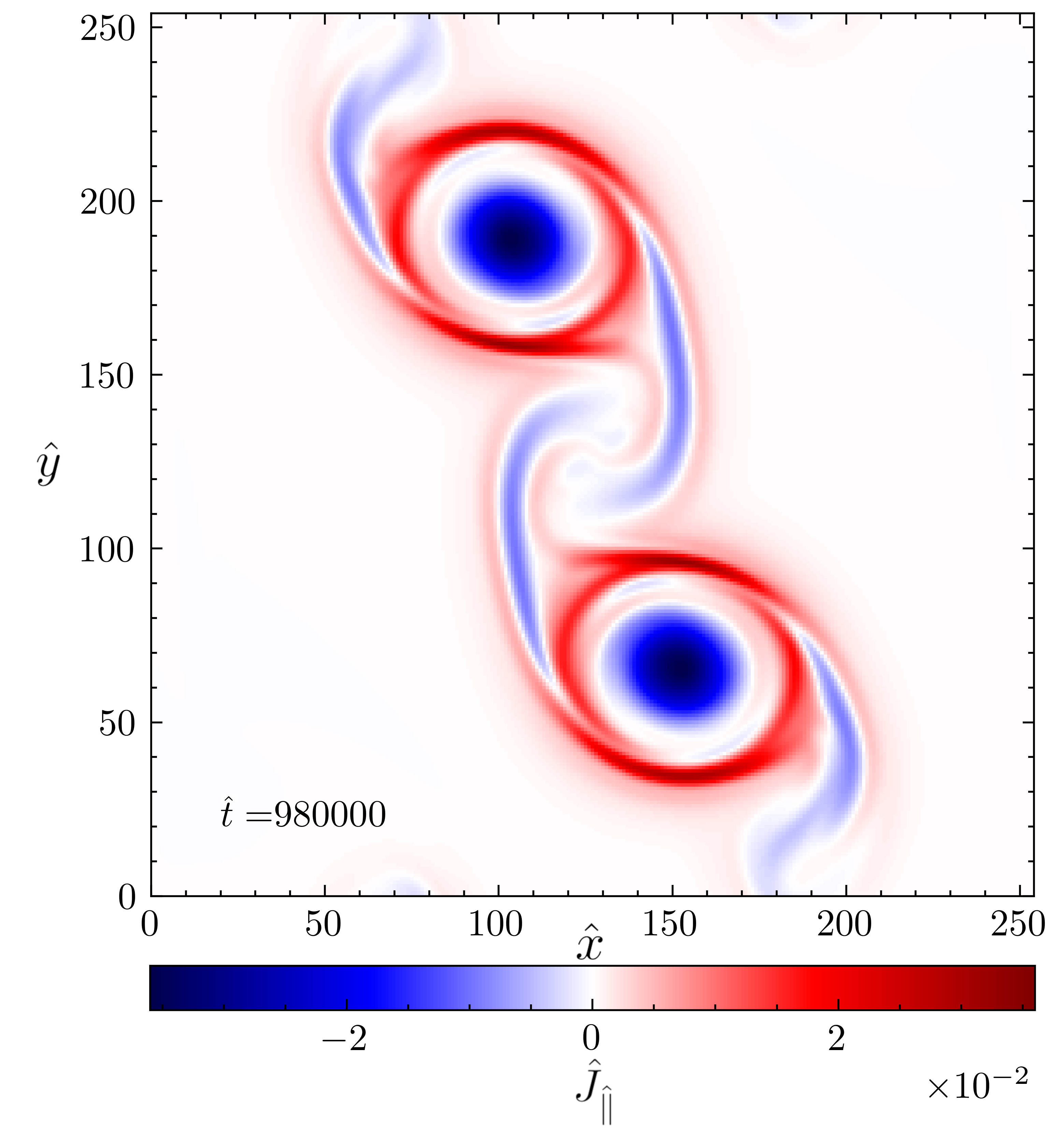}
  \caption{Parallel current $\hat{J}_{\hat{\parallel}}(\vbm{\hat{x}})$ for the interaction of two co-rotating magnetized vortices. Although the parallel currents attract, the vorticity formed between the islands prevents the merging process. }
  \label{magvort1}
\end{figure}

\begin{figure}[htp]
\centering
  \includegraphics[trim=0 2 0 2,clip,width=0.5\columnwidth]{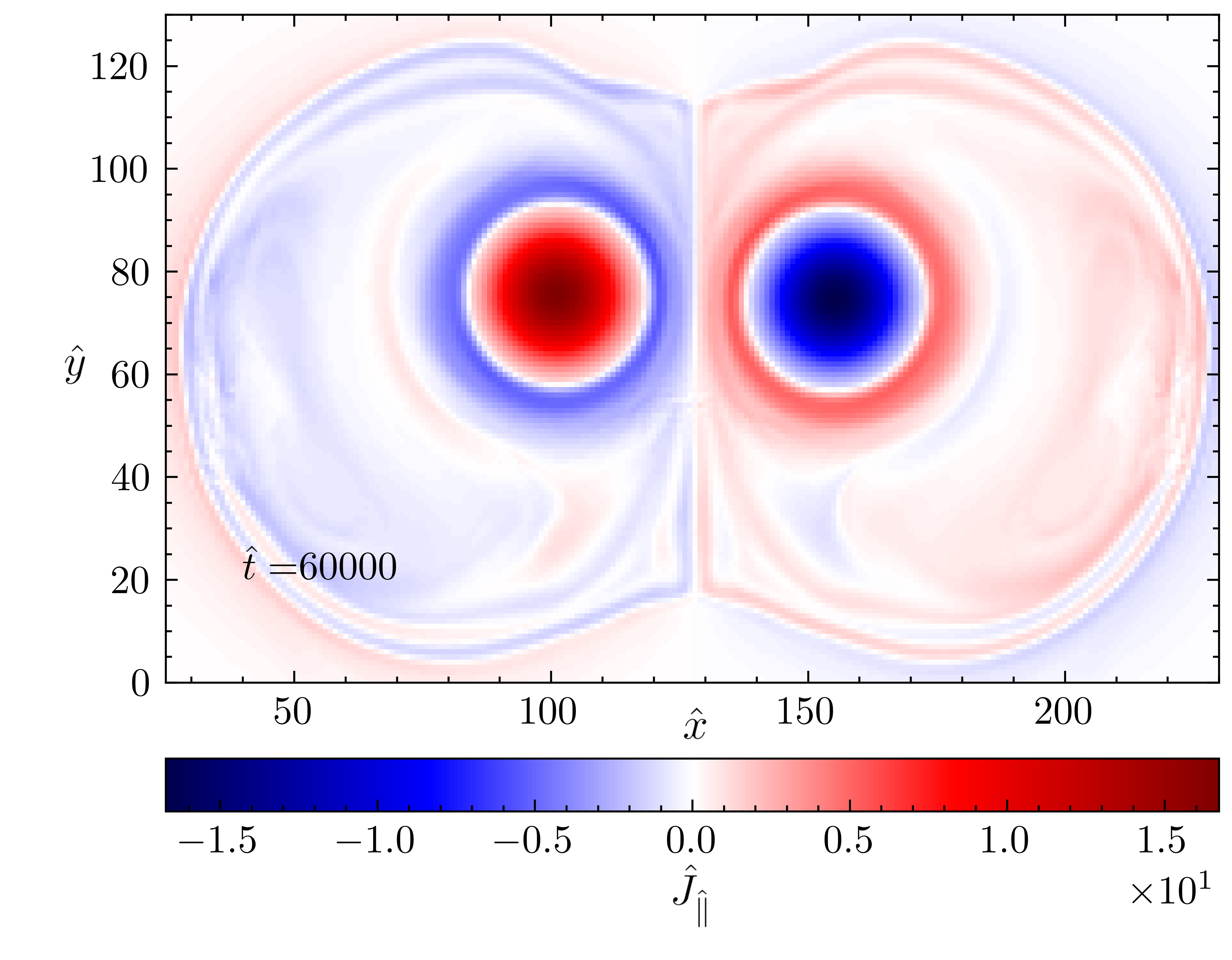}
  \caption{Parallel current $\hat{J}_{\hat{\parallel}}(\vbm{\hat{x}})$ for the interaction of two co-rotating magnetized vortices with a repelling magnetic configuration. The co-rotating vortices become encircled by an extensive vorticity layer that carries over to the parallel current and prevents vortex merging.}
  \label{magvort2}
\end{figure}

Combining several effects, the interaction of magnetic islands or magnetized vortices can be simulated by GREENY. In the simplistic picture, parallel currents attract each other and merge into a tube-like structure. On the other hand, co-rotating vortices, dependent on FLR-effects, can merge while counter rotating vortices will co-advect \cite{KendlVort}. When the same Gaussian profile is chosen for both, the initial vorticity and magnetic vector potential, one can study the interplay between repelling/attracting rotating currents. As shown in figures \ref{magvort1} and \ref{magvort2}, the aligned currents $\hat{J}_{\hat{\parallel}}(\vbm{\hat{x}})=\hat{n}_e(\vbm{\hat{x}})\hat{u}_e(\vbm{\hat{x}})+\hat{N}_i(\vbm{\hat{x}})\hat{U}_i(\vbm{\hat{x}})$ are attracting each other but the initial vorticity (co-rotating) forms an encircling vorticity $\hat{W}=\hat{\nabla}_{\hat{\perp}} \hat{\phi}$, delaying the merging process.

\section{Conclusion}
\label{conclu}
GREENY (Gyrofluid Reconnection
with Extended Electromagnetic Nonlinearity) was presented and proven to pass relevant verification tests such as solver tests and conservation laws. It contains several models and limits that allow a comparison of FLR effects on 2D magnetic reconnection in different settings. We were able to reproduce results in situations dominated by drift wave dynamics with strong gradients in the electric potential $\phi$, as well as magnetically driven reconnection processes. Further, we have found strong indication, that subgrid dissipation in form of a hyperviscosity, can alter the dynamics of a Harris-sheet reconnection event drastically. It appears like one can not provide a rule of thumb for choosing the hyperviscosity parameter $\hat{\nu}$, as the behaviour is too erratic. We suggest to choose the parameter small enough, so the influence are negligible on the energetic level, but no oscillatory behaviour and no granular structures are observed.
This work is intended as a reference publication for future use of the code and/or further developed versions. Future and updated versions, as well as documentation, will be available on \url{https://git.uibk.ac.at/c7441315/greeny}.\\

\textbf{Funding and Acknowledgement}
This work has been carried out within the framework of the EUROfusion Consortium, funded by the European Union via the
Euratom Research and Training Programme (Grant Agreement No 101052200 — EUROfusion). Views and opinions
expressed are however those of the author(s) only and do not necessarily reflect those of the European Union or the European
Commission. Neither the European Union nor the European Commission can be held responsible for them.\\

 The author thanks Bruce Scott and Matthias Wiesenberger (DTU) for valuable discussions.\\

\subsection{Author Contributions}
\textbf{F. F. Locker}: Data Curation (lead); Conceptualization (equal); Validation (equal); Software (equal); Visualization (equal); Writing - original draft; Methodology (equal); Formal analysis (lead); Project Administration (equal);
\textbf{M. Held}: Methodology (equal); Supervision (equal);
\textbf{T. M. Stocker Waldhuber}: Validation (supporting); Visualization (equal); Software (equal); Data Curation (support);
\textbf{A. Stürz}: Validation (equal); Visualization (equal); Software (supporting);
\textbf{M. Rinner}: Validation (supporting); Visualization (equal); Software (supporting);
\textbf{A. Kendl}: Supervision (equal) ; Conceptualization (equal); Writing – review \& editing (equal); Resources; Project Administration (equal);

\textbf{Declaration of competing interest}
The authors declare that they have no known competing financial 
interests or personal relationships that could have appeared to influence the work reported in this paper.

\textbf{Data availability}
The full code is availaible on \url{https://git.uibk.ac.at/c7441315/greeny} and the presented data can be made available after contacting the main author.




\bibliographystyle{elsarticle-num}
\bibliography{cpcbib}







\end{document}